\documentclass[twocolumn,showpacs,preprintnumbers,pra,aps,superscriptaddress,longbibliography,footinbib]{revtex4-1}

\usepackage{graphicx}
\usepackage{multirow}
\usepackage{amsmath}
\usepackage{mathtools}
\usepackage{amssymb}
\usepackage{soul}
\usepackage{verbatim}
\usepackage{bm}
\usepackage[colorlinks=true, linkcolor=blue, citecolor=gray]{hyperref}
\usepackage[table]{xcolor}
\usepackage{tikz}
\usepackage{xspace}
\usepackage{import}
\usepackage[caption=false]{subfig}
\usepackage{bbold}

\makeatletter
\newcommand\footnoteref[1]{\protected@xdef\@thefnmark{\ref{#1}}\@footnotemark}
\makeatother

\newcommand{\ket}[1]{\ensuremath{\left\lvert{#1}\right\rangle}}
\newcommand{\bra}[1]{\ensuremath{\left\langle{#1}\right\rvert}}

\newcommand{\pnorm}[1]{\lVert{#1}\rVert_\perp}

\newcommand{\tS}{\mathrm{S}} 
\newcommand{\tB}{\mathrm{B}} 
\newcommand{\tF}{\mathrm{F}} 
\newcommand{\tC}{\mathrm{C}} 
\newcommand{\tR}{\mathrm{R}} 

\newcommand{\hamsys}{H_{\mathrm{S}}}
\newcommand{\hamcoupling}{H_{\mathrm{C}}}

\DeclareMathOperator{\Tr}{Tr}

\begin{document}

\title{Quantum digital cooling}

\author{Stefano Polla}\email{polla@lorentz.leidenuniv.nl}
\affiliation{Instituut-Lorentz, Universiteit Leiden, P.O. Box 9506, 2300 RA Leiden, The Netherlands}
\author{Yaroslav Herasymenko}
\affiliation{Instituut-Lorentz, Universiteit Leiden, P.O. Box 9506, 2300 RA Leiden, The Netherlands}
\author{Thomas E. O'Brien}
\affiliation{Instituut-Lorentz, Universiteit Leiden, P.O. Box 9506, 2300 RA Leiden, The Netherlands}
\affiliation{Google Research, Venice, CA 90291, United States}

\begin{abstract}
	We introduce a method for digital preparation of ground states of simulated Hamiltonians, inspired by cooling in nature and adapted to leverage the capabilities of digital quantum hardware.
	The cold bath is simulated by a single ancillary qubit, which is reset periodically and coupled to the system non-perturbatively.
	Studying this cooling method on a 1-qubit system toy model, we optimize two cooling protocols based on weak-coupling and strong-coupling approaches.
	Extending the insight from the 1-qubit system model, we develop two scalable protocols for larger systems.
	The LogSweep protocol extends the weak-coupling approach by sweeping energies to resonantly match any targeted transition.
	We test LogSweep on the 1D tranverse-field Ising model, demonstrating approximate ground state preparation with an error that can be made polynomially small in the computation time for all three phases of the system.
	The BangBang protocol extends the strong-coupling approach, and exploits a heuristics for local Hamiltonians to maximise the probability of de-exciting system transitions in the shortest possible time.
	Although this protocol does not promise long-time convergence, it allows for a rapid cooling to an approximation of the ground state, making this protocol appealing for near-term demonstrations.
\end{abstract}

\maketitle

Ground state preparation is an essential algorithm in the quantum computing toolbox.
Any polynomial-time quantum algorithm can be mapped to the problem of estimating the ground state energy of an artificial Hamiltonian given an approximation to its ground state~\cite{wocjan2006several}, and without such additional input this problem is known to be QMA-hard for even 2-local Hamiltonians~\cite{kempe2006complexity}.
Digital quantum simulation of problems in materials science and chemistry, one of the `killer apps' of a quantum computer, is most often concerned with properties of ground states of the simulated systems~\cite{lloyd1996universal,reiher2017elucidating}, and many problems in optimization may be mapped to ground state finding problems~\cite{farhi2000quantum,farhi2014quantum}.
This has led to a wide range of schemes for digital ground state approximation, via adiabatic evolution~\cite{farhi2000quantum}, variational methods~\cite{peruzzo2014variational,farhi2014quantum,mcclean2016theory}, phase estimation~\cite{xu2014demon}, amplitude amplification \cite{poulin2009preparing,lemieux2021resource,lin2020near} and approximate imaginary time evolution and other Hamiltonian function techniques~\cite{mcardle2018variational,motta2019quantum,kyriienko2019quantum,ge2019faster}.
However, these algorithms suffer from large computational costs or approximation errors, making designing better schemes an active area of interest.

In nature, ground states are achieved by coupling to a large cold reservoir, which takes energy from the system in keeping with the second law of thermodynamics. 
Simulating an entire bath would require an impractically large quantum register, however it has long been suggested that this may be mimicked by coupling to a single qubit which may be reset to its ground state with sufficient frequency~\cite{lloyd1996universal}.
This idea has been since studied in digital quantum computing for the initialization of quantum devices~\cite{boykin2002algorithmic,kielpinski2000quantum} and as an inspiration of an algorithm based on resonant transitions and postselection \cite{wang2017quantum}. 
This idea was also explored in analog simulation settings, for the preparation of physical~\cite{popp2006ground} and artificial~\cite{raghunandan2019initialization,Metcalf2019} ground states.
However, cooling an artificial system in the digital quantum setting provides a set of unique challenges --- the system being studied may be completely different from the physical quantum hardware, and the digitized Hamiltonian may be only an approximation to the target of interest.
Furthermore, the periodic non-unitary reset may break the unitary evolution in short time-scale chunks which do not conserve energy,
implying that one may artificially reheat the system without clever protocol design.
This is of critical importance in near-term devices, where limited coherence times compete against the desire for slower cooling cycles.

In this work, we detail how one may prepare ground states of an artificial Hamiltonian on a digital quantum computer via quantum digital cooling (QDC).
We first present an analytic study of the cooling of a two-level system, from which two different approaches may be outlined to de-excite to the ground state whilst preventing reheating.
We investigate the behaviour of both methods in the digitized setting, and find they continue to be robust.
The protocols deriving from these two principles are tested in the one-qubit
black-box Hamiltonian setting, where the energy gap and matrix elements are unknown.
We extend these protocols to $N$-qubit systems, and investigate their ability to cool small-scale simulations of the transverse-field Ising model numerically.
Our LogSweep protocol, based on the weak-coupling approach, is demonstrated to converge to the ground state of all three phases of the transverse-field Ising model.
It further shows a relative energy error constant in the system size at a fixed protocol length for the weakly-coupled and critical phases of this model, which corresponds to a requirement to simulate time evolution for $O(N^2)$ and $O(N^3)$ Trotter steps respectively.
By contrast, the stong-coupling BangBang protocol shows the ability to prepare low-cost ground-state approximations of the same model in the paramagnetic and ferromagnetic regime, but seems to perform much worse close to the critical point, where the system spectrum shows a less-ordered structure.
The small number of calls to the system evolution operator needed to realize this protocol makes it attractive for near term application.

\section{Cooling a system with a single fridge qubit}

\begin{figure}[h!] 
	\centering{	
	\includegraphics[width=1\columnwidth]{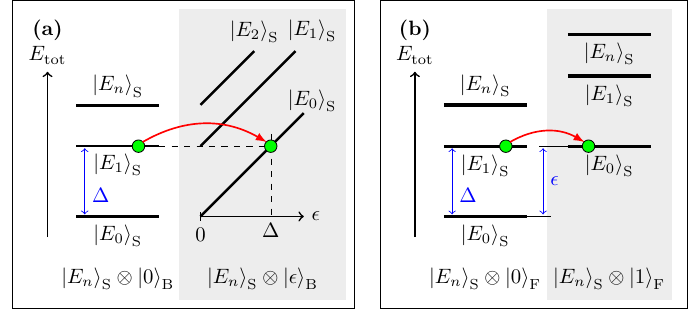} 
	}
	\caption{
		The de-excitation of the system transition $\ket{E_1}_\tS \to \ket{E_0}_\tS$ mediated by: \textbf{(a)} a continuous-spectrum natural bath B, where an excitation $\ket{\epsilon}_\tB$ at energy $\epsilon$ is produced, and \textbf{(b)} a single-qubit digital fridge F, which can be excited if $\epsilon = \Delta$.
	} 
	\label{fig:level-scheme}	
\end{figure}

In nature, gapped physical systems cool down to a state with high overlap to the ground state when interacting with a bath that is cold and large, under the condition of ergodicity.
By \emph{cold}, we mean that temperature $T_\tB$ of the bath is small compared to the ground state gap $\Delta_\tS$ of the system to be cooled: $k_\text{B}T_\tB \ll \Delta_\tS$ (with $k_{\text{B}}$ Boltzmann's constant).
By \emph{large}, we mean that the bath has a sufficiently large Hilbert space that the above condition is unchanged by the addition of the energy from the system.
By \emph{ergodic}, we mean the system must not have symmetries that prevent excitations to be transferred from the system to the bath, or that reduce the effective size of the accessible bath Hilbert space.
Given a system with Hamiltonian $\hamsys$ and eigenstates $\hamsys|E_j\rangle=E_j|E_j\rangle$, energy conservation implies that the bath must have states at energies $E_j-E_0$ to allow de-excitation of the eigenstates $E_j$.
This is typically achieved by considering a bath with a continuous or near-continuous low-energy spectrum [Fig.~\ref{fig:level-scheme}(a)].
However, the bath need not cool all states immediately to the ground state.
Instead, a bath typically absorbs single quanta of energy $\epsilon=E_i-E_f$ that correspond to local excitations of the system $|E_i\rangle \to |E_f\rangle$, at a rate given by Fermi's golden rule:
\begin{align}
&\frac{d P_{\mathrm{i}\rightarrow \mathrm{f}}}{dt}=\frac{2}{\hbar}\int_0^\infty d\epsilon\,|\langle E_f,\epsilon|\hamcoupling|E_i,0\rangle|^2\,\rho_B(\epsilon)\nonumber\\
&\hspace{1cm}\times\lim_{t\rightarrow\infty}\frac{\sin[(E_i-E_f-\epsilon)\,t]}{E_i-E_f-\epsilon}\label{eq:fgr1}\\
&=\frac{2\pi}{\hbar}\,|\langle E_f,\epsilon|\hamcoupling|E_i,0\rangle|^2\,\rho_B(E_i-E_f)\label{eq:fgr2},
\end{align}
where $\hamcoupling$ is the coupling between the system and the bath, and $\rho_B$ is the density of states of the bath 
\footnote{In the rest of the paper we assume $\hbar=1$}.
This requires the bath to be large enough to prevent reexcitation of these states as the system continues cooling.
In other words, the bath must have a large Hilbert space compared to the one of the system.
This ensures that, at equilibrium, most of the entropy is distributed in the bath.

To represent such a large bath with an ancillary register on a quantum device in order to cool a system register would be impractically costly.
In this work, we approximate the presence of a much larger bath B with a single ancilla qubit F [Fig.~\ref{fig:level-scheme}(b)], with bath Hamiltonian $H_\tB\to
H_\tF=\epsilon\,Z_\tF/2$.
This can be considered a simplified form of a collisional model~\cite{kretschmer2016collision} that does not allow for non-Markovian effects (that would be in our case unwanted).
The coupling between the bath and the system takes the form $\hamcoupling=\gamma X_\tF\otimes V_S/2$, where $\gamma$ is the coupling strength, and $V_S$ a \emph{coupling term} that acts on the system alone.
A key advantage of the digital approach is that we are free to choose $V_S$ as desired to optimize the cooling protocols.
The Hamiltonian of the entire system and bath then takes the form
\begin{equation}
H=\hamsys+H_\tF+\hamcoupling.\label{eq:full_ham}
\end{equation}
This has an immediate problem, as the bath can only absorb a single quantum of energy $\epsilon$, but we may circumvent this by periodically resetting the ancilla qubit to $\ket{0}$.
The non-unitary reset in effect extracts energy and entropy from the ancilla to a much larger external bath (the experimenter's environment). 
For this reason we call the ancilla qubit a `fridge' qubit (hence F).
The non-unitarity introduced in the process is necessary to dissipate entropy, allowing to prepare the ground state from an arbitrary starting state.
As the time between resets is finite, the $t\rightarrow\infty$ limit in Eq.~\eqref{eq:fgr1} is no longer justified and energy is no longer conserved.
This is both a blessing and a curse --- we need not precisely guess the energy gap $\Delta =E_i-E_f$ of the transition that we need to de-excite, but we run the risk of reheating the system at each cooling round.
Minimizing re-heating while maximizing the range of targeted de-excitations is key to the successful design of QDC protocols.
In a realistic experiment, qubit re-heating would be effectively increased by reset infidelity on the ancilla qubit, making the protocol less effective.

\section{De-exciting a single transition: the 1+1 model}

In order to design some basic protocols for QDC, we turn to a toy `1+1' model.
We take a single-qubit system with Hamiltonian $\hamsys=\Delta\,Z_\tS/2$, and couple it to a single fridge qubit with coupling term $V_\tS=X_\tS$.
Although this model is simple, it can for instance represent a pair of levels being targetted for cooling in a much larger quantum system. We will make use of this interpretation when extending these  cooling protocols in section~\ref{sec:Scalable_QDC}.

\subsection{Elementary approaches to digital cooling: strong and weak-coupling \label{sec:known_gap_cooling}}

Let us first assume $\Delta$ is known, in which case resonant cooling ($\epsilon=\Delta$) can be seen to be the most effective choice of $\epsilon$.
With this fixed, the transition probabilities after time $t$ may be calculated exactly to be
\begin{equation} \label{eq:transition-probabilities}
	P_{1\rightarrow 0}=\sin^2\left(\frac{\gamma}{2} t\right)
	, \,
	P_{0\rightarrow 1}=\frac{\gamma^2\sin^2(t\Omega)}{4\Omega^2},
\end{equation}
where $\Omega=\sqrt{\gamma^2/4+\epsilon^2}$.
We wish to maximise the cooling probability $P_{1\rightarrow 0}$ while minimizing the reheating probability $P_{0\rightarrow 1}$ by optimizing the remaining free parameters: the coupling strength $\gamma$ and the cooling time $t$.
To maximize the cooling rate $P_{1\rightarrow 0}=1$, we must set
\begin{equation}
	t=\pi \gamma^{-1}.\label{eq:coupling_time}
\end{equation}
We assume this constraint throughout this paper. This goes beyond the perturbative regime $\gamma t \ll 1$ in which Eq.~\eqref{eq:fgr1} is formulated.
However, we can take two very different approaches to minimize reheating, based on strong or weak coupling.
The weak-coupling approach is based on the observation that the off-resonant transition $P_{0\rightarrow 1}$ is bounded by $\gamma^2/4\Omega^2$.
As such, we may suppress reheating to an arbitrary level by choosing sufficiently small $\gamma$.
The time-cost for Hamiltonian simulation of $e^{iHt}$ scales at best linearly in $t$~\cite{Berry2007efficient}, so this implies one may obtain the ground state with failure probability $p$ in time $O(p^{-1})$, regardless of the initial state of the qubit.
The strong-coupling approach consists of tuning $\gamma$ so that $\Omega t=\pi$, which is achieved when 
\begin{equation}\gamma=\frac{2}{\sqrt{3}}\epsilon.\label{eq:strong_coupling}
\end{equation}
This fixes the reheating exactly to $0$, guaranteeing the qubit to be in the ground state in the shortest possible time, but at the cost of requiring fine-tuning.

\begin{figure}[h!]
	\includegraphics[width=\columnwidth]{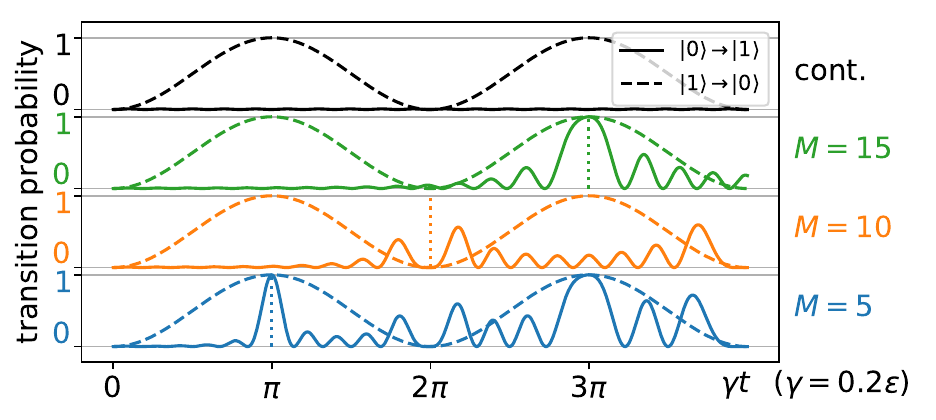}
	\caption{
Effects of Trotterization on cooling and reheating probabilities as a function of the coupling time $t$, for different numbers of Trotter steps $M$ per cooling cycle.
Vertical dotted lines indicate the $M$-th reheating oscillation, at which point the Trotter approximation fails.
	}
	\label{fig:trotter}
\end{figure}

Unlike in analog quantum simulation, digital devices cannot exactly implement the dynamics of the Hamiltonian in Eq.~\eqref{eq:full_ham}, and must approximate it digitally instead.
A common approach to such digitization is that of the Suzuki-Trotter expansion~\cite{Trotter,Suzuki}, which we now explore for the two cooling paradigms.
We apply the (second-order) expansion of the coupled system-bath evolution with Trotter number $M$,
\begin{align} \label{eq:trotter}
	e^{-i(\hamsys+H_\tF+\hamcoupling)\,t}
	\sim \left[e^{-i\hamcoupling\frac{t}{2 M}}e^{-i(\hamsys+H_\tF)\frac{t}{M}}e^{-i\hamcoupling\frac{t}{2 M}}\right]^M.
\end{align}
Note that, when we later approach larger systems, we will practically realize $e^{-i\hamsys t/ M}$ as a single second-order Trotter step, effectively implementing a second-order Trotter simulation of the coupled Hamiltonian $\hamsys+H_\tF+\hamcoupling$ with trotter number $M$.
If we restrict to the subspace containing the states involved in the cooling transition \mbox{$\ket{10}_{\tS\tF}\to\ket{01}_{\tS\tF}$}, at resonant cooling we have \mbox{$\hamsys + H_\tF \propto \mathbb{1}$} (specifically, in this model $\ket{01}$ and $\ket{10}$ generate a zero-eigenvalue subspace of $\hamsys + H_\tF$).
Thus, the Trotterized evolution behaves exactly like the continuous one with regards to the cooling transition.
We study reheating probabilities as a function of t for different values of $M$ in the weak-coupling regime. 
We observe (Fig.~\ref{fig:trotter}) that the digitized evolution approximates well the behavior of the continuum limit whenever $t\Omega/\pi \lesssim M$ (i.e. for the first $M$ Rabi oscillations with pulse $\Omega$).
For longer times $t\Omega\pi \gtrsim M$, the second-order Trotter approximation fails, leading to reheating rates far larger than in the continuum limit.
This allows us to define a practical choice of $M$ to avoid reheating due to digitization --- we require
\begin{equation}
	M \gg \sqrt{1+\epsilon^2/\gamma^2},\label{eq:Trotter_step}
\end{equation}
which sets the working point $t=\pi\gamma^{-1}$ before the $M/2$ Rabi oscillation.
However, in the strong-coupling case $t\,\Omega/\pi = \sqrt{3}$, which implies that a single step is sufficient. 
Indeed, digitized cooling with probability $1$ and no reheating can be realized by a \emph{bang-bang} approach (inspired by similar approach in variational methods~\cite{Yang2016,Bapat2018}). This consists in defining the evolution as in Eq.~\eqref{eq:trotter} with $M=1$, as long as the coupling strength is adjusted to $\gamma = 2 \epsilon$.
With this choice, the digitized evolution implements resonant Ramsey interference on the cooling transition $\ket{10}_{\tS\tF}\to\ket{01}_{\tS\tF}$ and anti-resonant Ramsey interference on the reheating transition $\ket{00}_{\tS\tF}\to\ket{11}_{\tS\tF}$.

\begin{figure}[h] 
	\includegraphics[width=1\columnwidth]{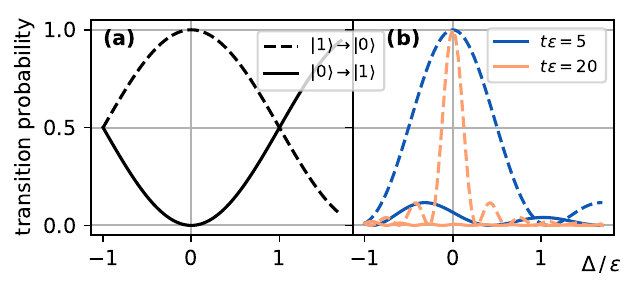}
	\vspace{-2em}
	\caption{
		Effect of fridge-system detuning $\delta = \Delta - \epsilon$ on the cooling (dashed lines) and reheating (solid lines) probabilities for \textbf{(a)} the bang-bang cooling approach, and \textbf{(b)} the weak-coupling cooling approach, where colors indicate different couping strengths.
	} 
	\label{fig:detuning}
\end{figure}

A key difference between the two approaches to digital cooling is in their behavior off-resonance, i.e.~when the energy gap is mistargetted or not precisely known.
For the bang-bang approach, detuning reduces the cooling efficiency while symmetrically boosting reheating [Fig.~\ref{fig:detuning}(a)].
The wide resonance peak around zero detuning makes this approach ideal to quickly cool transitions which energy is known up to a small error.
In the weak-coupling approach the resonance peak becomes sharper and the reheating gets more suppressed as the coupling is made smaller [Fig.~\ref{fig:detuning}(b)], approaching the energy conservation limit.
Detuning makes cooling inefficient, but thanks to the low reheating probability this weak-coupling cooling can be iterated while changing $\epsilon$ to try to match the transition energy, without destroying the cooling effect.

\subsection{Common symmetries and the coupling alternation method \label{sec:common_symmetries}}

In large systems of interest, we do not know the Hamiltonian's eigenstates, making it more challenging to couple between them.
This is required for cooling, as can be seen by the direct dependence of the cooling rate $\frac{dP_{i\rightarrow f}}{dt}$ on the overlap $|\langle E_f,\epsilon|H_\mathrm{C}|E_i,0\rangle|^2$ (Eq.\ref{eq:fgr2}).
This overlap dependence will prohibit cooling if the system Hamiltonian $H_\tS$ and the coupling operator $V_\tS$ share a common symmetry $S$ (i.e., $\left[S,H_\tS\right]=\left[S,V_\tS\right]=0$).
When this is the case, the Hamiltonian may be simultaneously diagonalized with $H_\tS$, and a state with some overlap to any eigenspace of $S$ that does not contain the ground state cannot be cooled to the ground state by coupling with $V_\tS$.
Note that this is a strictly stronger condition than just requiring $[H_\tS,V_\tS]\neq 0$.
A simple solution is to alternate over a set of couplings $\{V^i_\tS\}$ as we cool.
Then, any symmetry $S$ of $H$ need commute with \emph{each} $V^i_\tS$ in order to guarantee that a state starting with overlap in a high-energy eigenspace will remain there.
Sets of coupling terms $\{V^i_\tS\}$ on $N$ qubits need only be size $O(N)$ to break commutation with all non-trivial operators (for example, the set of all single-qubit Pauli operators), so although symmetries need to be taken into account, they will not destroy the feasibility of QDC protocols.

\begin{figure}[h]
	\includegraphics[width=\columnwidth]{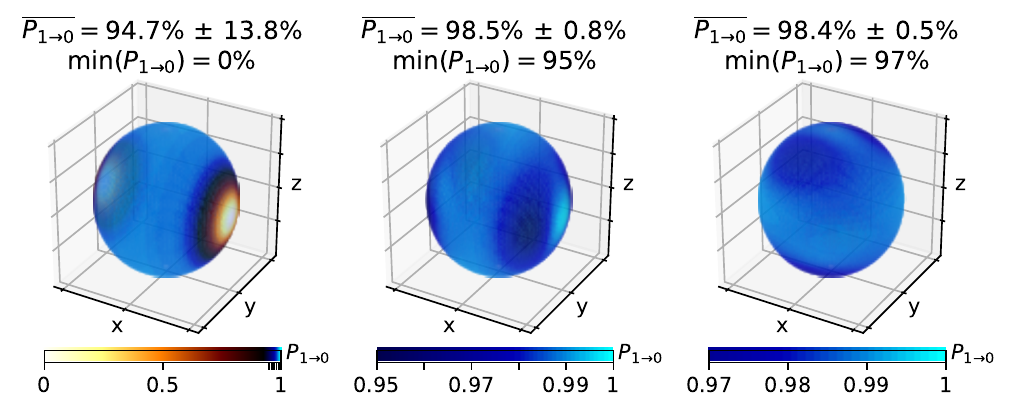}
	\caption{ \label{fig:XYZ-spheres}
		Probabilities $P_{1\to0}$ of transitioning from $\ket1$ to $\ket0$ after three iterations of the weak-coupling ($t\,\epsilon = 10$) cooling procedure, with coupling potentials $V^i_\tS=X, X, X$ (left), $V^i_\tS=X, Y, X$ (center), $V^i_\tS=X, Y, Z$ (right),
		on a system qubit with Hamiltonian $H_\tS=h\,\vec{n}\cdot\vec{\sigma}$ and known energy splitting $h$.
		The orientation of the unit vector $\vec{n}$ is represented on spherical surfaces. 
		The average, standard deviation and minimum of $P_{1\to0}$ are shown above each panel.
	}
\end{figure}

This issue may be demonstrated on the protoype $1+1$ qubit model by considering the system Hamiltonian $H_\tS=h\,\vec{n}\cdot\vec{\sigma}$, where $\vec{n}$ is a $3$-dimensional unit vector (so $H_{\tS}$ points in a random direction on the Bloch sphere), $2h$ is a fixed energy splitting, and $\sigma$ is the vector of Pauli-matrices.
For a fixed coupling operator $V_\tS$, there is a risk that $[H_\tS, V_\tS]\approx 0$. 
When this is the case, the off-diagonal elements of $V_\tS$ in the system eigenbasis are zero, preventing cooling. 
We may prevent this by alternating between different coupling terms during the cooling protocol, such that no non-trivial Hamiltonian can commute with all such coupling terms.
This may be achieved for the $1+1$ model by iterating over $V^i_\tS\in\{X_\tS,Y_\tS,Z_\tS\}$, or alternatively over $V^i_\tS\in\{X_\tS,Z_\tS\}$.
The effectiveness of this scheme is studied in Fig.~\ref{fig:XYZ-spheres} for resonant coupling. We see the probability $P_{1\rightarrow 0}$ of successful cooling of the weak coupling approach ($t\,\epsilon=10$) increased to $\min(P_{1\rightarrow 0})=97\%$ for all choices of $\vec{n}$ when iterating over $V^i_\tS = X_\tS,Y_\tS,Z_\tS$, and above $95\%$ when iterating $V^i_\tS = X_\tS,Z_\tS,X_{\tS}$, compared to the possibility for complete cooling failure [$\min(P_{1\rightarrow 0})=0$] when $V^i_\tS$ is held constant.
Similar results are seen for off-resonant coupling.

\section{Scalable QDC protocols \label{sec:Scalable_QDC}}

We now look to use the insight obtained for cooling in the 1+1 toy model to develop QDC schemes for larger systems.
The sub-additivity of entropy places a rough lower bound on the number of cooling steps required to cool an $N$-qubit system.
This limits the entropy $\Delta S_\tS$ that the system can transfer to the fridge qubit before the non-unitary reset to $\Delta S_\tS \geq -\Delta S_B \geq -1$ bit.
If we demand the ability to cool an arbitrary state, a QDC protocol must also be able to cool the maximally-mixed state, which has entropy $S_\tS = N$.
We then require $N$ repetitions of an optimal coupling-and-reset step to reach the pure ground state (which has entropy $S_\tS=0$).
This can be obtained in the simple example of cooling $N$ non-interacting qubits with known energies, by simply repeating the protocols of the $1+1$ model.
However, this cannot be generalised to arbitrary strongly-correlated systems, as cooling is complicated by irregular and unknown energy gaps and coupling terms between eigenstates.
This is to be expected, as preparing ground states of arbitrary Hamiltonians is a known QMA-hard problem~\cite{kempe2006complexity}.
However, as cooling is a physically-motivated process, we hope QDC to be able to achieve polynomial scalings for systems of physical interest, i.e. models of systems that are found in low-temperature equilibrium states in nature.
We focus for the rest of the work on exploring this thesis.

In the rest of this text, we introduce two scalable QDC protocols for $N$-qubit systems: the strong-coupling-based BangBang protocol and the weak-coupling-based LogSweep protocol.
These extend and generalize the two approaches we established for the 1+1 toy model of section~\ref{sec:known_gap_cooling}.
Each protocol iterates over a sequence of \emph{cooling steps}, each of which consists of coupling the fridge qubit to part of the system for a short time evolution, and then resetting the fridge qubit to its ground state.
The protocols differ in the choice of coupling strengths $\gamma_i$, coupling terms $V_S^i$ and fridge energies $\epsilon_i$ at each $i$-th cooling step.
[The coupling time for each cooling step is fixed by Eq.~\eqref{eq:coupling_time}].

\subsection{The BangBang protocol}\label{sec:bangbang}
We now develop a protocol to extend the strong-coupling approach (Eq.~\ref{eq:strong_coupling}) to a larger system.
This motiviation is in line with recently proposed bang-bang approaches to adiabatic state preparation~\cite{Yang2016,Bapat2018}, which are known to outperform initial theoretical expectations stemming from a naive Trotter error estimate.
We are thus optimistic that this 'BangBang' protocol may provide a low-cost, near-term method for QDC.
However, such a protocol needs to associate each fridge-system coupling with a single fridge energy, that should match the transitions that this coupling triggers.
An appropriate choice of fridge energy can be estimated via a perturbation theory approximation.
To derive this approximation, we note that the rate of a transition between eigenstates $\ket{E_i}\to\ket{E_j}$ depends on the matrix element of the coupling $V_S$:
\begin{equation}
	V_{(ij)} := \bra{E_{i}} V_S \ket{E_{j}} = \frac{\bra{E_{i}} [\hamsys,V_S] \ket{E_{j}}}{E_{i}-E_{j}}.
\end{equation}
If $V_S$ is local and bounded, $[\hamsys,V_S]$ is as well, so the matrix element $V_{(ij)}$ will be bounded proportionally to $(E_{i}-E_{j})^{-1}$.
The matrix element is additionally bounded by $\lVert V \rVert$; this second bound becomes dominant when $E_i-E_j / \lVert V \rVert $ falls below the maximum off-diagonal element of $[H,V]$ in any basis, which we define with the notation $\pnorm{[H, V]}$:
\begin{equation} \label{eq:transition-norm}
	 \lVert O \rVert_\perp
	 = \max_{\langle\phi\vert\psi\rangle = 0}\!\left| \bra{\phi} O \ket{\psi} \right|
	 = \max_{\ket{\Phi}, \ket{\Psi}}\!\frac{\bra{\Phi} O \ket{\Phi} - \bra{\Psi} O \ket{\Psi}}{2},
\end{equation}
where O is Hermitian and the maxima are taken over all possible states $\ket{\psi}, \ket{\phi}$ and $\ket{\Psi}, \ket{\Phi}$. A simple proof is given in Appendix \ref{app:norm-proof}.
We use this energy scale to set the fridge energy: 
\begin{equation}
\epsilon_i = \lVert[V_\tS^i, H_\tS]\rVert_\perp \label{eq:bang_bang_fridge_energy}
\end{equation} 
for any coupling potential $V_\tS^i$.
This way, the maximum-energy transitions accessible by $V_\tS$ are on resonance, while smaller energy ones (which are less important for cooling) still have a higher probability of cooling than of reheating [see Fig.~\ref{fig:detuning}(a)].
This defines the BangBang protocol: we iterate over coupling to each qubit, with $\epsilon_i$ fixed by Eq.~\eqref{eq:bang_bang_fridge_energy}.
As this protocol does not attempt to suppress reheating, we choose a single coupling $V_S=Y_n$ for the $n$-th qubit, instead of iterating over $V_S=X_n,Y_n,Z_n$ (as was suggested in Sec.\ref{sec:common_symmetries}). In general, the best choice of $V_\tS$ will depend on the system that we want to cool, and the couplings should be picked to enable as many transitions as possible.
We repeat the coupling to each qubit $R$ times, resulting in a total of $RN$ cooling steps.
Each cooling step contains two first-order Trotter steps simulating $e^{-i H_\tC t/2}$ (of depth $d^{(1)}_{H_\tC}$), a single second-order Trotter step for $e^{-i H_\tS t}$ (of depth $d^{(2)}_{H_\tS}$), and a reset gate, resulting in a total circuit depth
\begin{equation} \label{eq:bangbang-depth}
	d_\text{BangBang} = R N (2 d^{(1)}_{H_\tC} + d^{(2)}_{H_\tS} +1).
\end{equation}

\begin{figure}[h] 
	\centering{
	\includegraphics[width=.95\columnwidth]{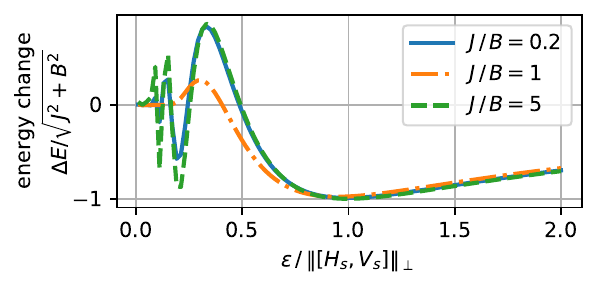}}
	\caption{
		Change in energy expectation value for the application of a single cooling step to the maximally mixed state of a $N=8$ qubit transverse field Ising chain Eq.~\eqref{eq:ising}, depending on the fridge energy $\epsilon$.
		The coupling potential is $V_\tS=Y_3$, the Pauli $Y$ on the third qubit.
		The relation $B^2+J^2=1$ fixes the energy scale.
	} 
	\label{fig:energyguess}
\end{figure}

To test the BangBang protocol, we study the cooling of a $N$-qubit transverse-field Ising chain 
\begin{equation} \label{eq:ising}
	H_\tS = \sum_{i=0}^N B X_i + \sum_{i=0}^{N-1} J Z_i Z_{i+1} ,
\end{equation}
where $B$ represents the transverse magnetic field Zeeman splitting and $J$ is the Ising coupling strength.
The relative coupling strength $J/B$ dictates whether the system is in the paramagnetic ($J/B\ll1$), ferromagnetic ($J/B\gg1$), or critical ($J/B\sim 1$) phases.
This ability to simply tune between three phases of matter with significantly different physical properties make the TFIM a good benchmark model to investigate the ability of different QDC schemes in various scenarios.

We first demonstrate that our choice for the fridge energy Eq.~\eqref{eq:bang_bang_fridge_energy} is appropriate.
In Fig.~\ref{fig:energyguess}, we plot the effect of a single cooling step on the maximally-mixed state.
We observe that cooling is maximized for fridge energies around the point defined by Eq.~\eqref{eq:bang_bang_fridge_energy}, for all phases of the TFIM.
We find this behaviour to hold for all other (local) choices of coupling potential $V_S$ used in this work, as predicted.

\begin{figure}[h] 
	\includegraphics[width=\columnwidth]{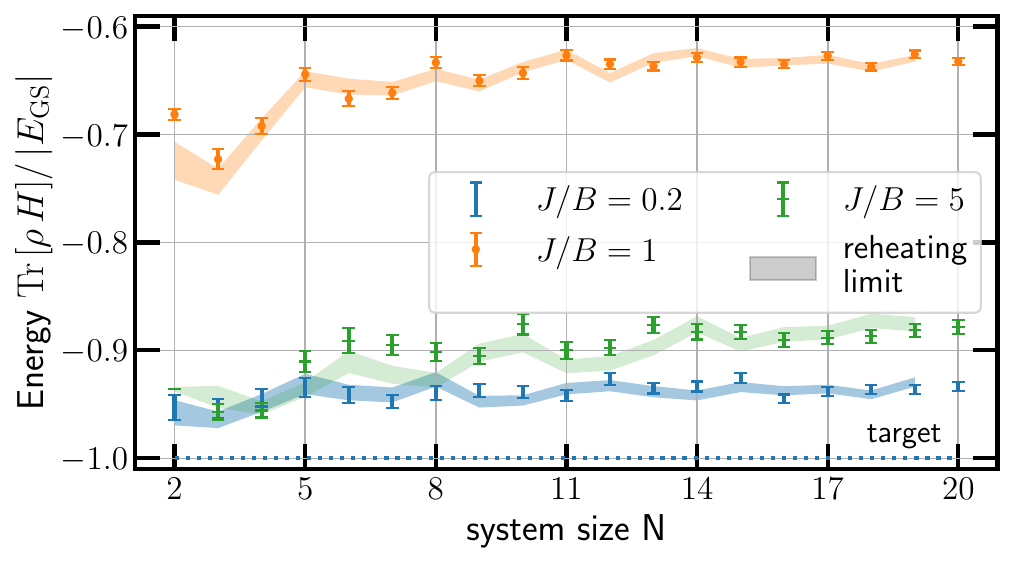}
	\caption{
		Performance of the BangBang protocol as a function of the system size $N$ for the three different phases of the transverse-field Ising model (detailed in legend). The coupling potentials are $V_\tS^i = Y_i$. 
		Dots correspond to result when the protocol is applied to the maximally-mixed state, shaded regions corresponds to result when protocol is applied to the true ground state (which gives a bound on protocol re-heating).
		Data generated by Trotterized wave-function simulations of the protocol, and random sampling of the initial mixed state and of nonunitary operations (details in App.~\ref{app:methods}). 
		All points are run with 200 samples, and average results are plotted with the sampling error.
	} 
	\label{fig:bangbang}
\end{figure}

We next turn to the ability of the BangBang protocol to prepare an approximation $\rho$ of the ground state, starting from a maximally-mixed (i.e. infinite temperature) initial state. 
We benchmark by the final state with two figures of merit: the ground state fidelity
\begin{equation} \label{eq:fidelity}
	F=\Tr\big[\,|E_0\rangle\langle E_0| \,\rho\,\big],
\end{equation} 
and the energy relative to the ground state energy $Tr[H_\tS \rho]/|E_\text{GS}|$.
This last property is local in local system, and represents an energy density in TFIM.
To verify convergence, we compare cooling results to a \emph{reheating limit}, obtained by running the protocol with the ground state as initial state.
We observe that all TFIM phases converge after $R \approx N$ repetitions (with the weakly-coupled phase system converging already at the first repetition).
In Fig.~\ref{fig:bangbang} we plot the energy density of the cooled state, as well as the reheating limit, as a function of the number of sites in the system.
This shows that convergence is indeed achieved for $R=N$ independently on the system phase and size, and that the final energy density stays approximately constant, without showing any other trend.
The BangBang protocol achieves a final energy density close to 90\% and 95\% of $\pnorm{H_\tS}$ for the ferromagnetic and paramagnetic regime respectively, while performing significantly worse in the critical regime.
This is to be expected, as in this regime the spectrum is no longer banded, and excitation energies are not as uniform as in the paramagnetic or ferromagnetic regimes.
Following Eq.~\eqref{eq:bangbang-depth}, the protocol's circuit depth is $7NR$ for a gate-set containing single- and double-qubit rotations (and the reset gate).
Given the low cost of the protocol, we suggest that this is of particular interest for near-term experiments, and may be further refined by other cooling protocols, or methods such as quantum phase estimation, in the long term.

\subsection{The LogSweep protocol \label{sec:LogSweep}}

Refrigeration at weak-coupling suppresses reheating, but only allows for the cooling of transitions within a narrow energy band [as shown in Fig.~\ref{fig:detuning}(a)].
We may take advantage of this in a larger system, where a wide range of energy gaps are present, by sweeping the fridge energy $\epsilon_k$ from high to low as we iterate over cooling steps.
(As low-energy transitions are more susceptible to re-heating than high-energy transitions, this will in general be more efficient than sweeping from low to high.)

To construct a full protocol, we further need to fix the set of fridge energies $\epsilon_k$ and linewidths $\delta_k=t_k^{-1}=\pi\,\gamma_k$ that we plan to use for each cooling step. We will be guided by two principles. First, the target band of fridge energies $(E_{\min},E_{\max})$ should be tightly covered by the cooling lines $\epsilon_k\pm\delta_k$. Second, the reheating should be suppressed to a certain degree throughout the protocol. 
As by Eq.~\eqref{eq:transition-probabilities} the reheating suppression depends on $\gamma_k/\epsilon_k$, we fix this value to a small constant throughout the protocol (i.e. we choose $\gamma_k \propto \epsilon_k$).
Thus we define the LogSweep protocol, where the fridge energy $\epsilon_k$ is sweeped over $(E_{\min},E_{\max})$ in a logarithmic gradation.
Specifically, given the \emph{gradation number} $K$, we index each cooling step $k=1,\ldots,K$, and we define
\begin{equation} \label{eq:logsweepepsilon}
	\epsilon_k = E_\text{min}^{ \frac{k-1}{K-1} } E_\text{max}^{ 1 - \frac{k-1}{K-1} },
\end{equation}
and choose $\delta_k$ to fix $\epsilon_{k+1}+\delta_{k+1}/\zeta = \epsilon_k - \delta_k/\zeta$, with $\zeta$ a constant (potentially dependent on $K$).
In App.~\ref{app:asymptotics}, we prove that such a scheme will cool a single transition in the range $(E_{\min},E_{\max})$ with probability $1$ as $K\rightarrow\infty$, and in App.~\ref{app:energy_opt} we demonstrate that the logarithmic gradation is optimal for such a scheme for a choice of $\zeta(K) \sim \log(K)$.
To make sure all system excitations have a chance of being dissipated, we further iterate the couplings $V_S$ over a set of local couplings  $\{V_S^i\}$  throughout the system:  for the considered spin systems we choose $\{V_\tS^i\}\equiv\{X_n,Y_n,Z_n\}$ for each qubit $n$ (see Sec.\ref{sec:common_symmetries}), for a total of $3NK$ cooling steps.
The number of Trotter steps $M_k$ for each cooling step $k$ is chosen to prevent re-heating.
This follows Eq.~\ref{eq:Trotter_step}, but as transition energies between system eigenstates may be as large as the Hamiltonian spread $2\pnorm{H_\tS}$, we set
\begin{equation}
	M_k = 2 \sqrt{1 + \frac{2\pnorm{H_\tS}^2}{\gamma_k^2}}.
\end{equation}
The choice of the fridge energy range $[E_\text{min}, E_\text{max}]$ will generally depend on heuristics on the system.
$E_\text{max}$ should be greater or equal than the largest energy of the transitions that we are able to de-excite with the chosen couplings $V_\tS$ (for local Hamiltonians we can estimate this with the techniques described in \ref{sec:bangbang}).
For ground state cooling, $E_\text{min}$ should be close to the system ground state gap $\Delta_\text{GS}$, as no transition with an energy lower than $\Delta_\text{GS}$ can lead from an excited state to the ground state.

\begin{figure}[h] 
	\centering{
	\includegraphics[width=\columnwidth]{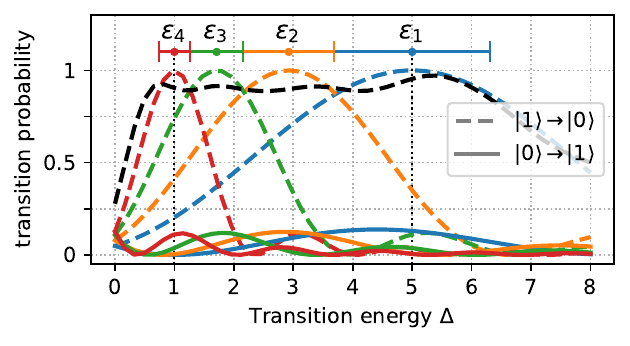}}
	\caption{
		Choices of energies $\epsilon_k$ and linewidths $\delta_k$ (bars at the top of the graph showing $\epsilon_k\pm\delta_k$) for a $K=4$ LogSweep protocol applied to the model introduced in section~\ref{sec:known_gap_cooling} with an unknown $\Delta\in \left(E_\text{min}=1, E_\text{max}=5\right)$.
		Colored lines show cooling (dashed) and reheating (solid lines) probabilities for each $j$-th step alone, the dashed black line shows the cooling probability after sequential application of the 4 steps.
	} 
	\label{fig:logsweep_1p1}
\end{figure}

We first test the LogSweep protocol as applied to the 1+1 model defined in Sec.~\ref{sec:known_gap_cooling}, with the system gap $\Delta$ now taking an unknown value between $E_\text{min}$ and $E_\text{max}$ (Fig.~\ref{fig:logsweep_1p1}).
At each step $k=1,\ldots,K$ we want to maximise cooling of transitions $\Delta\sim\epsilon_k$, while minimizing reheating of previously-cooled transitions $\Delta\sim \epsilon_{k'}$, $k'<k$.
As demonstrated by the black dashed curve in Fig.~\ref{fig:logsweep_1p1}, when $E_\text{max}/E_\text{min}=5$ this can be achieved well with only $K\approx E_{\max}/E_{\min}$ steps.
Note that, to maintain a constant relative linewidth (and thus constant maximum reheating per step), we should scale $K \sim E_\text{max}/E_\text{min}$.
This implies $K\rightarrow \infty$ as $E_{\min}\rightarrow 0$, in line with the third law of thermodynamics.

\begin{figure}[h]
	\includegraphics[width=\columnwidth]{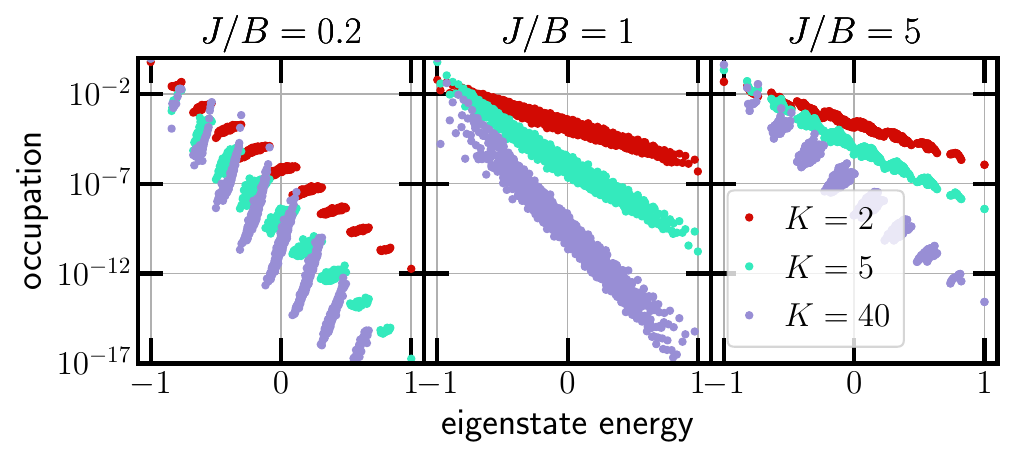}
	\caption{\label{fig:banding}Effect of banding on single LogSweep iterations. A maximally mixed state in the three different phases of the 7-qubit TFIM spin chain is evolved by the LogSweep protocol for three different values of $K$. We plot the distribution of the result here over the system's eigenstates (indexed by energy), at three different values of $K$. We see that while the critical system demonstrates an approximate thermal or exponential distribution, the weak and strongly-coupled systems demonstrate an inversion in the population of the system within each band, which increases with $K$. Data generated by continuous-evolution density-matrix simulation (details in App.~\ref{app:methods}).}
\end{figure}

In a larger system, the situation is more complex than in the model above.
Instead of a single transition from the excited state $\ket{E_1}\rightarrow \ket{E_0}$ which occurs with unit probability when $\epsilon=\Delta= E_1 - E_0$, our system may transition to one of many eigenstates $\ket{E_j}$, to each with a transition probability $A_{i,j}$ (assuming a initial state $\ket{E_i}$). As there are many possible target states, the maximum transition probability might be very small ($\max_j A_{i,j}\ll1$).
If we restrict to a single transition $|E_i\rangle\rightarrow |E_j\rangle$ with the above reduced cooling rate, one may show that the LogSweep protocol still cools that transition with unit probability as $K\rightarrow \infty$, albeit at a rate that scales exponentially in $A_{i,j}$.
Luckily, we do not need to ensure any specific transition occurs, instead we may cool sequentially
\begin{equation}\label{eq:sequential}
	\ket{E_i} \rightarrow \ket{E_{j_0}} \rightarrow \ket{E_{j_1}} \rightarrow \ldots \rightarrow \ket{E_0},
\end{equation}
with a growing number of possible cooling paths as the system grows and the transition probabilities spread over more eigenstates.
A good choice of the fridge energy interval $[E_\text{min}, E_\text{max}]$ and of the coupling potentials $\{V_\tS^i\}$ allows all eigenstates to be connected to the ground state by sequences of transitions $|E_{j_l}\rangle\rightarrow |E_{j_{l+1}}\rangle$ that have unit probability of being de-excited for $K\to\infty$.
However, a single transition probability approaches 1 only over the entire LogSweep protocol.
In particular, if the transition $|E_{j_l}\rangle\rightarrow|E_{j_{l+1}}\rangle$ during step $k^*$ of the protocol corresponds to an energy loss $E_{j_l}-E_{j_{l+1}} \gg \epsilon_{k^*}$, this transition will be off-resonance for the entire remaining duration of the protocol (as $\epsilon_k<\epsilon_{k^*}$ for $k>k^*$), making it unlikely to occur.
This can cause convergence issues especially when cooling systems with banded spectra.
For such systems , as we set $E_{\min} \approx \Delta_\text{GS}$ as detailed above, there may be a point $k^*$ in the protocol after which $\epsilon_k$ will become smaller than the average interband gap, but never as small as the spread of a single band.
After this point, states at the bottom of a band might transition to states in the lower band, but states at the top of each band never have any resonant transitions to lower energy states, thus becoming absorbing states.
This effect is clearly shown in Fig.~\ref{fig:banding}, representing the LogSweep-cooled states of the transverse-field Ising model in different regimes.
We start with the maximally-mixed state, and plot the resultant distribution over the eigenstate energies.
In the banded regimes (side panels), we observe that the distribution of energies in any given band is tilted towards the higher-energy states in that band (i.e. the aforementioned absorbing states), by some orders of magnitude.
This dead-ends ultimately hinder sequential cooling, and prevent the LogSweep cooling from converging to the same state independently on the initial state.
The effect worsens as $K$ is increased, as transition linewidths $\delta_k$ become smaller making off-resonant transitions less and less probable.
This issue can be fixed in practice by using an initial state with fewer high-energy excitation (e.g. a classical approximation of a low-energy state).
We solve the issue in principle, by constructing an \emph{iterative LogSweep} protocols, where the LogSweep cooling is repeated with growing $K$.
The early, lower-cost iterations cool the highest energy excitations, while the larger $K$ iterations grant vanishing reheating, and probabilities approaching unity for the cooling transitions allowed by symmetries.
Thus, adding iterations with larger and larger $K$, will make the whole protocol converge to the system ground state (unless symmetries forbid all paths from some states to the ground state).
Note that this adjustment is not required for systems with a continuous spectra (i.e.~critical systems), as in such a system there will be on-resonance transitions for any state with an energy $E_{\min}$ or more above the ground state.

\begin{figure}[h]
	\includegraphics[width=\columnwidth]{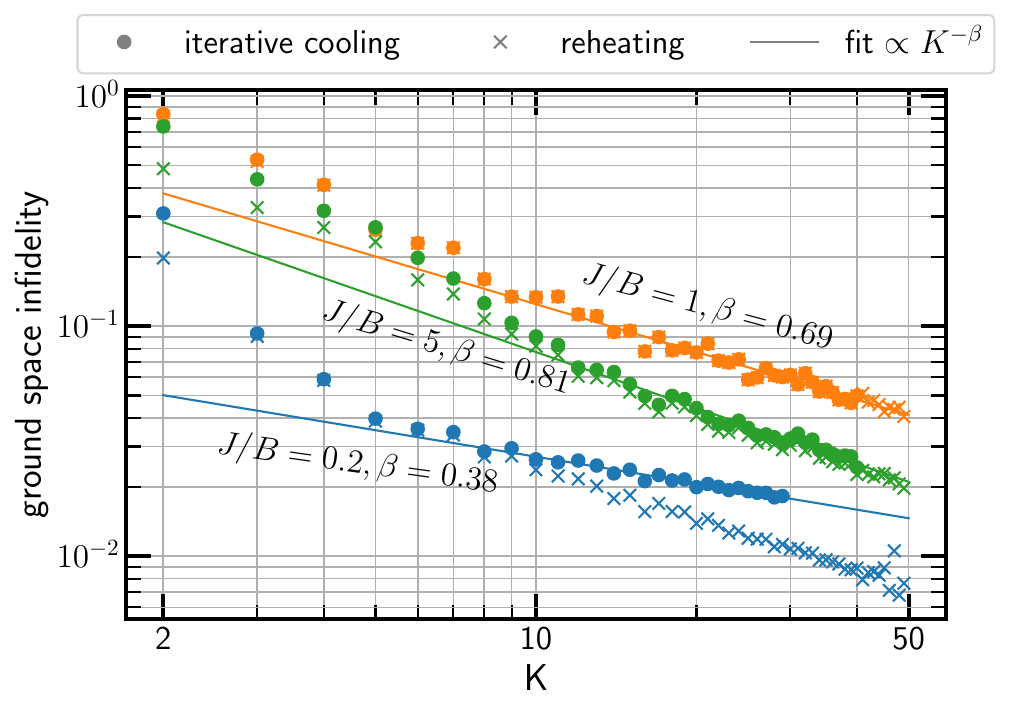}
	\caption{\label{fig:K_infidelity}Convergence of the LogSweep protocol to the ground state as a function of the gradation number $K$, starting from the maximally mixed state (dots) and the ground state (crosses), for three phases of the transverse-field Ising model (detailed in legend). Data was generated by deterministic density-matrix simulations of the iterative LogSweep protocol, with second-order Trotter Hamiltonian simulation (details in App.~\ref{app:methods}).}
\end{figure}

We now investigate the performance of the (iterative) LogSweep protocol on different phases of the transverse-field Ising model.
In Fig.~\ref{fig:K_infidelity}, we plot the ground state infidelity of the prepared state $\rho$ [$1-F$ with F as in Eq.~\eqref{eq:fidelity}], as a function of $K$.
The protocol consists in $K-1$ sweeps of a LogSweep QDC protocol, each sweep having gradation number $g_l=2,\ldots,K$.
The Hamiltonian simulation is performed by second-order Trotter approximation.
We investigate the protocol effect on two initial states $\rho_0$: the maximally-mixed state $\rho_0 = \mathbb{1}/2^n$ to check for cooling capabilities (dots), and the ground state $\rho_0 = \ket{E_0}\bra{E_0}$ (crosses) to show the lower bound originated by reheating.
We observe polynomial convergence to the ground state in all three phases of the model, attaining an infidelity of $\varepsilon = 1-F$ in approximately $K \sim O(\varepsilon^{-1/\beta})$ energy gradation steps for $\beta\approx0.4\,\text{-}\,0.8$.
Additionally, we verify that the protocol converges to the reheating limit for the critical and strongly-coupled regimes. 
In the weakly-coupled regime instead, although the cooling is far more efficient because of the local nature of the system excitations, the reheating bound is not saturated. We attribute this to the very small linewidths $\{\delta_k\}$, consequence of the well-defined transition energies, together with the strong banding of the system spectrum.

The number of Trotter steps for a single iteration of the LogSweep protocol with gradation number $g_l$ on a system of $N$ spins with Hamiltonian $H_\tS$ scales as $O\!\left(\pnorm{H_\tS} \Delta_\text{GS}^{-1}  N g_l^2 \log(g_l)^{-1}\right)$.
Thus, the iterative implementation required to deal with the banded cases needs a total number of Trotter steps
\begin{equation}
	M_\text{tot} \sim O\!\left(\pnorm{H_\tS} \Delta_\text{GS}^{-1}  N K^3 \log(K)^{-1}\right)
	\label{eq:LogSweep_circuit_scaling}
\end{equation}
The gate complexity required to attain an error (infidelity) $\varepsilon$ for the models studied scales thus as $O(\varepsilon^{-3})$ - $O(\varepsilon^{-8})$.

\begin{figure}[h]
	\includegraphics[width=\columnwidth]{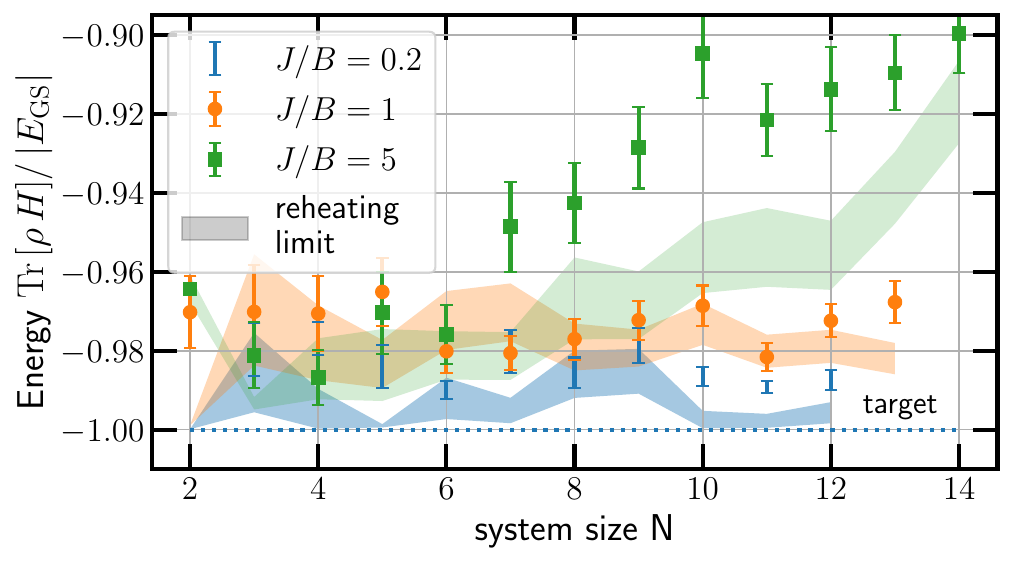}
	\caption{\label{fig:LSscaling}
		Performance of the LogSweep protocol as a function of the system size for the three different phases of the transverse-field Ising model (detailed in legend), with fixed $K=5$. Dots correspond to result when protocol is applied to the maximally-mixed state, shaded region corresponds to result when protocol is applied to the true ground state (which gives a bound on protocol re-heating).
		Data generated by Trotterized wave-function simulations of the protocol, and random sampling of the initial mixed state and of nonunitary operations (details in App.~\ref{app:methods}). 
		All points are run with 100 samples, and average results are plotted with the sampling error.
	}%
\end{figure}

We next investigate the scaling of the LogSweep protocol as a function of the system size.
In Fig.~\ref{fig:LSscaling} we plot the relative error in the ground state energy as a function of the system size for a single (not iterated) LogSweep with gradation number $K=5$.
We see a constant error in the ground state energy as a function of the system size for the weakly-coupled and critical systems. Thus, here we expect no need to scale $K$ with $N$ for the protocol to be accurate.
Let us also note that the gap in these two cases shrinks as $\Delta_\text{GS}/\lVert H \rVert \sim N^{-1}$ and $\Delta_\text{GS}/\lVert H \rVert\sim N^{-2}$ respectively. Using the above arguments and the estimate \eqref{eq:LogSweep_circuit_scaling}, one can find how the circuit length (in terms of time evolution steps), required to obtain a constant energy error, scales with $N$. We obtain $O(N^2)$ for the weakly-coupled and $O(N^3)$ for the critical case.
From this analysis, we expect that the QDC protocol may be asymptotically competitive with methods such as adiabatic state preparation, whose runtime naively scales as $O(1/\Delta^2_\text{GS})$ ~\cite{reiher2017elucidating, farhi2000quantum}.
In the strongly-correlated phase, we do not see such positive results; the energy error increases with the system size, though the relative error remains beneath $10\%$ for up to $14$ spins.
This may be explained by the relative growth of the extension of excitations within the strongly correlated phase, while cooling is performed with local couplings.
Due to the error in the simulation, we are unable to reliably extract an estimate of the computational cost in the same way as for the critical and weakly-coupled systems.
Future work may explore whether this error may be improved on by adjusting the form of the coupling terms $\{V_\tS^i\}$ based on heuristics on the considered system.

\section{Conclusion}
	
In this paper, we investigated how cooling can be simulated on a digital quantum computer, and demonstrated that this can be exploited for the design of scalable algorithms for preparing ground states of $N$-qubit systems.
We identified how one can meet many of the fundamental challenges that the digital approach to cooling raises and use the leverage offered exclusively by digital quantum hardware, namely the freedom of choice in the coupling strength and fridge energy.
We laid out a general approach of simulating a cold bath with a single ancilla qubit, which is iteratively coupled to various locations in the system and reset periodically to extract entropy and energy.
We studied how to digitize the system-fridge coupling simulation without causing additional reheating, and how to avoid symmetries which produce non-ergodic behavior that hinders cooling.
By tuning coupling parameters beyond the perturbative regime described by Fermi's golden rule, efficient cooling of targeted transitions can be realized.
Following these principles we proposed two protocols for preparing approximate ground states of $N$-qubit systems --- the BangBang protocol and the LogSweep protocol.
We studied numerically how these protocols perform on the three phases of the 1D transverse-field Ising model.
We found that the BangBang protocol quickly cools the system near to the ground state in the paramagnetic and in the ferromagnetic regime, but has difficulty in the critical regime.
The LogSweep protocol is observed to cool all three phases to the ground state at a polynomial cost in the overlap error.
In the weakly-coupled and critical phases, the LogSweep protocol further demonstrates a constant energy error as a function of the system size (for fixed gradation number), making it a competitive state preparation method.

Thanks to the low number of steps required, we believe the BangBang protocol has the potential of finding a near-term application, specially if integrated with projective or variational methods to improve its performance.
The requirement of precise Hamiltonian simulation and multiple cooling steps makes the LogSweep algorithm in its current form unsuitable for near-term implementation. 
Nevertheless, the scaling arguments discussed at the end of section \ref{sec:LogSweep} show that our algorithm can be competitive with other non-NISQ methods such as projective quantum phase estimation (QPE) and adiabatic state preparation.
QPE in its standard form requires multiple ancillas and an initial state with a finite ground state overlap, while as our has no requirement on the initial state.
Adiabatic state preparation requires an integrable Hamiltonian which can be adiabatically connected to the required Hamiltonian, and requires time scaling as  $O(1/\Delta^2_\text{min})$ ~\cite{reiher2017elucidating, farhi2000quantum} where $\Delta_\text{min}<\Delta_\text{GS}$ is the minimum gap along the adiabatic path.

The introduction of quantum digital cooling opens future research directions related to the characterization of proposed protocols, their optimization, and their extension beyond ground state preparation.
A study of the effect of noise on currently proposed QDC protocols, and the optimization of such protocols for noise resilience, are in order to establish their applicability on near-term devices.
Applying QDC to more complex physical systems, in areas such as quantum spin liquids, many-body localization and quantum chemistry, would bring new challenges to the protocol construction.
A thorough study of the role in the cooling process played by the symmetries and locality of coupling could lead to the design of more optimized protocols.
Furthermore, various extensions to the QDC protocols proposed in this work can be suggested.
In a parallelized version of QDC, the use of multiple fridge qubits coupled to various locations in the system might allow to trade space complexity for time complexity.
A variationally-optimized QDC protocol might be devised, that can efficiently prepare a state in the ground state manifold of some Hamiltonian starting from an arbitrary initial state --- differently from the variational quantum eigensolver \cite{peruzzo2014variational} which requires the preparation of a fiducial state at every iteration.
The principles of QDC might inspire a new class of efficient non-unitary quantum algorithms, where non-unitary operations are mediated by a single ancillary qubit, with possible application e.g.~in the simulation of open quantum system dynamics.

One application of particular future interest for QDC protocols is in the preparation of Gibbs thermal states, which are useful e.g.~for semi-definite programming \cite{Brandao2017quantum}.
This seems especially promising given the near-thermal distribution in Fig.~\ref{fig:banding} of the critical system under the evolution of the LogSweep scheme.
However, it is as of yet unclear how to overcome the finite width of the distribution, and how well these protocols behave in the banded case (or for more general systems).
Adjustment of the LogSweep protocol to produce robust thermal state preparation schemes is an obvious target for future research.

\begin{acknowledgments}
The authors would like to thank C.W.J.~Beenakker for support and advice during this project, B.~Tarasinksi, X.~Bonet Monroig, M.~Pacholski, and S.~Yalouz for useful criticism on the manuscript. This research was funded by the Netherlands Organization for Scientific Research (NWO/OCW) under the NanoFront and StartImpuls programs, and by Shell~Global~Solutions~BV.
\end{acknowledgments}


%


\appendix

\section{Proof of Eq.~\eqref{eq:transition-norm}} \label{app:norm-proof}
To prove Eq.~\eqref{eq:transition-norm} we first show that
\begin{equation} \label{eq:ineq1}
	\!\left| \bra{\phi} O \ket{\psi} \right|
	\leq 
	\max_{\ket{\Phi}, \ket{\Psi}}\!\frac{\bra{\Phi} O \ket{\Phi} - \bra{\Psi} O \ket{\Psi}}{2},
\end{equation}
for all $\ket{\psi}, \ket{\phi}:\langle\psi|\phi\rangle = 0$.
We can assume without loss of generality $\bra{\phi} O \ket{\psi}$ is real and nonnegative (if it's not, we can multiply one state by an irrelevant global phase), and drop the absolute value.
As $\langle\psi|\phi\rangle = 0$ we can define the states $\ket{\pm} = \frac{\ket{\phi} \pm \ket{\psi}}{\sqrt{2}}$ we can then write
\begin{align*} 
	\bra{\phi} O \ket{\psi}
	=&
	\frac{1}{2} ( \bra{\phi} O \ket{\psi} + \bra{\psi} O \ket{\phi})
	\\ 
	=&
	\frac{\bra{+} O \ket{+} - \bra{-} O \ket{-}}{2}
\end{align*}
immediately proving Eq.~\eqref{eq:ineq1}.
The opposite inequality is proven by noticing that the $\ket{\Psi}$ and $\ket{\Phi}$ that maximize the right of Eq.~\eqref{eq:transition-norm} have to be eigenvalues (by the variational principle). With these, we can redefine the states $\ket{\pm} = \frac{\ket{\Phi} \pm \ket{\Psi}}{\sqrt{2}}$ which are also granted to be orthogonal, thus
\begin{align*} 
	\frac{\bra{\Phi} O \ket{\Phi} - \bra{\Psi} O \ket{\Psi}}{2}
	=&
	\Re[\bra{+} O \ket{-}]
	\\ 
	\leq &
	|\bra{+} O \ket{-}|
	\\ 
	\leq &
	\max_{\langle\phi\vert\psi\rangle = 0}\!\left| \bra{\phi} O \ket{\psi} \right|
\end{align*}
which combined with Eq.~\eqref{eq:ineq1} proves Eq.~\eqref{eq:transition-norm}.

\section{Asymptotic reheating and cooling probabilities for QDC protocols}\label{app:asymptotics}
Let us consider a two-state subsystem of a larger Hilbert space with a gap energy $E$, evolving under a QDC protocol on the $k$th step via a coupling term that does not mix the $\{|01\rangle,|10\rangle\}$ and $\{|00\rangle,|11\rangle\}$ subspaces (where the second index denotes the fridge).
Under this assumption, the evolution of the system within this space is a Markov process.
Following the main text, let the fridge energy on the $k$th step be $\epsilon_k$, the coupling strength be $\gamma_k$, and the time evolved for in the cooling protocol $t_k$.
Additionally, let the spacing of the fridge energies to be 
$$
	(\epsilon_k-\epsilon_{k+1}) 
	= \zeta (\delta_k + \delta_{k+1})
	= \frac{\alpha}{2} (\gamma_k+\gamma_{k_{+1}}),
$$
for some $K$-dependent $\alpha=\alpha(K) =\frac{2}{\pi \zeta(K)}$.
We may calculate the transition matrix for the Markov process, $p^{(k)}(E)$ (defined by $p_{a,b}^{(k)}(E)=P(|a\rangle\rightarrow |b\rangle)$ in a single cooling step) as
\begin{equation}
	p^{(k)}(E)=\left(\begin{array}{cc}1-\sin^2(\Omega_kt_k/2)\frac{\gamma_k^2}{\Omega_k^2} & \sin^2(\omega_kt_k/2)\frac{\gamma_k^2}{\omega_k^2}\\
		\sin^2(\Omega_kt_k/2)\frac{\gamma^2_k}{\Omega_k^2} & 1-\sin^2(\omega_kt_k/2)\frac{\gamma_k^2}{\omega_k^2}
	\end{array}\right),\label{eq:markov_process}
\end{equation}
where
\begin{align}
	\omega_k=\sqrt{(E-\epsilon_k)^2+\gamma_k^2}\\
	\Omega_k=\sqrt{(E+\epsilon_k)^2+\gamma_k^2}.
\end{align}
Assuming no additional cooling or heating to the rest of the system during the protocol, the transition matrix for the $k_0\rightarrow k_1$ block takes the form
\begin{equation}
	P_{k_0,k_1}(E)=\prod_{k=k_0}^{k_1}p^{(k)}(E),
\end{equation}
and the transition matrix for the entire process may be written $P(E)=P_{1,K}(E)$.

Exact analytic evaluation of this expression in the large $K$ limit is quite difficult. Instead, we aim for a conservative estimate, bounding the final cooling probability $p_c=\left[P(E)\right]_{01}$ from below. For this, given the energy $E$, we first lower bound the `initial' cooling around the resonant step $k_c$, i.e. such $k_c$ that $|\epsilon_{k_c}-E|$ is minimal. Then we give an upper bound on reheating during the following protocol steps $k=k_c,..K$. Given the estimated cooling probability $p^{(k_c)}_c$ and reheating probability $p^{(k_c;K)}_{rh}$, we can obtain a lower bound for $p_c$:
\begin{equation}
p_c> (1-p^{(k_c;K)}_{rh})p^{(k_c)}_c \label{eq:final_cooling}
\end{equation}
The value of $p^{(k_c)}_c$ can be conservatively estimated from the formula:
\begin{align}
1-p^{(k_c)}_c&< \prod^K_{k=1} (1-\sin^2(\omega_{k}t_{k}/2)\frac{\gamma_{k}^2}{\omega_{k}^2})\\
&< \prod_{k, \frac{|E-\epsilon_{k}|}{\gamma_k}<1 }((E-\epsilon_{k})^2/\gamma^2_{k}),\label{eq:cooling_k_c}
\end{align}
where the second line follows from the inequality $\sin(\frac{\pi\sqrt{1+x^2}}{2})/(1+x^2)\geq\min(0, 1-x^2)$ applied to each term in the product. In the perfect resonance scenario, $|E-\epsilon_{k_c}|=0$ and the cooling probability is exactly $1$. The worst case scenario is when $E$ is right between the two neighbouring $\epsilon_k$'s, thus $|E-\epsilon_{k_c}|=\frac{\alpha}{2}\gamma_k$. In this case, we can calculate the logarithm of \eqref{eq:cooling_k_c} in the leading order of $K^{-1},~\alpha$:
\begin{align}
2\sum^{k^{(+)}_c}_{k=k^{(-)}_c}\log\left|\frac{\epsilon_k-E}{\gamma_k}\right|&= 2\int_{\epsilon^{(-)}}^{\epsilon^{(+)}}\log\left|\frac{\epsilon-E}{\gamma(\epsilon)}\right|d\epsilon\frac{dk}{d\epsilon}\label{eq:cooling_k_c_worst_step_1}\\
&=\frac{2}{\alpha}\int_{\epsilon^{(-)}}^{\epsilon^{(+)}}\log\left|\frac{\epsilon-E}{\gamma(\epsilon)}\right|\frac{d\epsilon}{\gamma(\epsilon)}. \label{eq:cooling_k_c_worst_step_2}
\end{align}

Here, we used the fact that $\gamma\alpha$ defines energy spacing (and so $\frac{d\epsilon}{dk}=\alpha\gamma(\epsilon)$), and introduced summation limits $k^{(\pm)}_c$, $\epsilon^{(\pm)}$ as the points where $\frac{\epsilon-E}{\gamma}=\pm1$. As this implies scaling $\epsilon^{(\pm)}=E+O(\gamma)$, \eqref{eq:cooling_k_c_worst_step_2} should scale as $O(1/\alpha)$.
The calculation can be completed for the LogSweep gradation $\epsilon_k,~\gamma_k$, which implies $\epsilon'_k\propto\gamma(\epsilon)\propto\epsilon$.
In particular, if $x=\frac{\epsilon-E}{\gamma}$ then $dx=\frac{Ed\epsilon}{\epsilon\gamma}=\frac{d\epsilon}{\gamma}(1+O(1/K))$, and we have:
\begin{align}
	\frac{2}{\alpha}\int_{\epsilon^{(-)}}^{\epsilon^{(+)}}\log\left|\frac{\epsilon-E}{\gamma(\epsilon)}\right|\frac{d\epsilon}{\gamma(\epsilon)}=\frac{4}{\alpha}\int^1_0\log x~dx=-\frac{4}{\alpha}\label{eq:cooling_k_c_worst}.
\end{align}
Substituting into Eq.~\ref{eq:cooling_k_c}, we find the initial cooling probability bounded by 
\begin{equation}
p^{(k_c)}_c\gtrsim1-\exp(-4/\alpha(K)).
\end{equation}

The reheating accumulated between steps $k_c$ and $K$, $p_{rh}^{(k_c;K)}$, can be upper bounded as:
\begin{equation}
p^{(k_c;K)}_{rh}\leq 1- \prod^K_{k=k_c}\left(1-\sin^2(\Omega_kt_k/2)\frac{\gamma_k^2}{\Omega_k^2}\right)
\label{eq:reheating_k_c_K}
\end{equation}
The product in Eq. \eqref{eq:reheating_k_c_K} can be further estimated as:
\begin{align}
\prod^K_{k=k_c}&\left(1-\sin^2(\Omega_kt_k/2)\frac{\gamma_k^2}{\Omega_k^2}\right)\geq\prod^K_{k=k_c}\left(1-\frac{\gamma_k^2}{\Omega_k^2}\right)\\
&\geq\prod^K_{k=k_c}\left(1-\frac{\gamma_k^2}{(E+\epsilon_k)^2}\right)\\
&\simeq\exp\left(-\sum^K_{k=k_c}\frac{\gamma_k^2}{(E+\epsilon_k)^2}\right),\label{eq:reheating_exponential}
\end{align}
where in the last line we assumed that $\gamma_k\ll E+\epsilon_k$ for all $k$. As we are most concerned about the large $K$ asymptotics of the total cooling probability, let us now analyze how the expression \eqref{eq:reheating_exponential} behaves in this limit. Since $\gamma^2_k$ scales as $O(1/K^2)$ and we have $K$ terms in the sum, we generally expect $O(1/K)$ scaling for the sum. Such scaling would imply a rapidly vanishing reheating for a large-$K$ protocol. In the specific case of the LogSweep protocol, to the leading order in $1/K$ one indeed obtains:
\begin{align}
& p^{(k_c;K)}_{rh}\lesssim\sum^K_{k=k_c}\frac{\gamma_k^2}{(E+\epsilon_k)^2}\approx\frac{1}{\alpha(K)}\int^E_{E_{\min}} \dfrac{\gamma(\epsilon)}{(E+\epsilon)^2}d\epsilon\\
&\approx\dfrac{\log\frac{E_{\max}}{E_{\min}}}{\alpha^2(K)K}(\frac{1}{2}-\frac{E}{E+E_{\min}}+\log(\frac{2E}{E+E_{\min}}))\\
&\equiv\dfrac{\tR(E_{\min},E_{\max},E)}{\alpha^2(K)K}.\label{eq:reheating_asymptotics}
\end{align}
Here, we used Eq. \eqref{eq:logsweepepsilon} and the fact that $\alpha(K)\gamma_k$ defines energy spacing $|\epsilon_{k+1}-\epsilon_k|$. 
Finally, combining Eqq. \eqref{eq:final_cooling} - \eqref{eq:reheating_asymptotics}, we obtain an asymptotic lower bound to the final cooling probability:
\begin{align}
p_c=\left(1-\exp\left(-\frac{4}{\alpha(K)}\right) \right)\cdot\left(1-\frac{\tR(E_{\min},E_{\max},E)}{\alpha^2(K) K}\right).
\end{align}
This estimate implies $p_c\rightarrow 1$ for large $K$, provided that both $e^{-4\alpha^{-1}(K)}\rightarrow0$ and $\frac{1}{K\alpha^2(K)}\rightarrow0$.

To ensure that the infidelity is minimized and thus $\alpha(K)$ is optimal, we solve the extremum condition $\partial_\alpha(e^{-4\alpha^{-1}}+\frac{R}{\alpha^2K})=0$ for $\alpha$. The solution can be expressed in terms of the product logarithm function $W$, $\alpha(K)=4~W^{-1}\left(8K/R\right)$. For large K, at the leading order we obtain simply: $\alpha(K)=4~\log^{-1}\left(8K/R\right)$. The infidelity then scales down almost linearly with K: $1-p_c=\frac{\log^2(8K/R)}{16K}$. This asymptotically optimal $\alpha(K)$ yields the choice $\zeta(K)=\frac{1}{2\pi}\log(8K/R)$, which we use in all our simulations.

\section{Optimizing energy spacing in LogSweep protocol}\label{app:energy_opt}
In Sec. \ref{sec:LogSweep}, we argued that the energy spacing of the LogSweep protocol is optimal for the protocol precision for a $K$-step protocol. This was based on the reheating estimate taken from the cooling step $k_c$ only. One may ask, if this persists when one includes the total reheating into account. In the large $K$ limit, we can use the estimate \eqref{eq:reheating_exponential} for this check. Fixing the constraint $\gamma_k=\frac{|\epsilon_{k+1}-\epsilon_k|}{\alpha}$, we proceed by means of variational calculus:
\begin{align}
\dfrac{\delta}{\delta \epsilon_k}\sum^K_{k=k_c}\frac{\gamma_k^2}{(E+\epsilon_k)^2}=0\\
\Rightarrow\dfrac{\delta}{\delta \epsilon(k)}\int^{K}_{k_c}\frac{(\epsilon'(k))^2}{(E+\epsilon(k))^2}dk=0\\
\Rightarrow\epsilon''(k)\cdot(E+\epsilon(k))=(\epsilon'(k))^2. \label{eq:var_sweepepsilon}
\end{align}
The solution to Eq. \eqref{eq:var_sweepepsilon} that satisfies boundary conditions $\epsilon(k_c)=E$, $\epsilon(K)=E_{\min}$, is as follows:
\begin{align}
\epsilon_k=(2E)^{\frac{K-k}{K-k_c}}(E+E_{\min})^{\frac{k-K}{K-k_c}+1}-E. \label{eq:var_sweepepsilon_solution}
\end{align}
This shows that the logarithmic character of the optimal spacing persists when we consider total reheating (cf. Eq. \eqref{eq:logsweepepsilon}). However, we cannot directly use the embelished result \eqref{eq:var_sweepepsilon_solution} for our cooling protocol. That is because this formula uses the targeted energy $E$ as a reference, whereas we are targetting a continuum of energies. Therefore, we keep using the simpler and more practical formula Eq. \eqref{eq:logsweepepsilon} for the LogSweep protocol.

\section{Cooling rate for LogSweep protocol in a large system}
In a large system, the above analysis is complicated by the presence of multiple transitions from every energy level. We now give a simplified analysis that focuses on a pair of states $|E_i\rangle$, $|E_j\rangle$, in a spirit similar to Appendix~\ref{app:asymptotics}. This means we formulate the protocol as a Markov process equivalent to (Eq.~\ref{eq:markov_process}), where the transitions to levels other than $i$ and $j$ are ignored. Note that in the perturbative limit, this is a good approximation of the actual Markov process as restricted onto the subspace $\ket{E_i}, \ket{E_j}$. Specifically, even though we ignore the indirect transitions between $i$ and $j$ via other levels, this is justified at the first order of pertubation theory.
Unlike in the $1+1$ model however, the transitions here are imperfect. If our total coupling has strength $\gamma$ (i.e. $\|H_C\|=2^N\gamma$), following the analysis in Sec.~\ref{sec:bangbang} the coupling between states $|E_i\rangle$ and $|E_j\rangle$ will take the form $\gamma \sqrt{A_{i,j}}$ with $\sqrt{A_{i,j}}$ scaling down as $O((E_i-E_j)^{-2})$.
This has the effect of scaling both the cooling and re-heating rates by $A_{i,j}$, recasting the Markov process (Eq.~\ref{eq:markov_process}) as

\begin{equation}
	p^{(k)}_{i,j}=\left(\begin{array}{cc}1-A_{i,j}\sin^2(\frac{\Omega_kt_k}{2})\frac{\gamma_k^2}{\Omega_k^2} & A_{i,j}\sin^2(\frac{\omega_kt_k}{2})\frac{\gamma_k^2}{\omega_k^2}\\
		A_{i,j}\sin^2(\frac{\Omega_kt_k}{2})\frac{\gamma^2_k}{\Omega_k^2} & 1-A_{i,j}\sin^2(\frac{\omega_kt_k}{2})\frac{\gamma_k^2}{\omega_k^2}
	\end{array}\right).\nonumber
\end{equation}
As this only reduces both the heating and cooling rates, our claim that reheating in the LogSweep protocol tends to $0$ as $K\rightarrow \infty$ still holds.
However, we need to repeat the analysis of App.~\ref{app:asymptotics} to bound the cooling rate $p_c^{(k_c)}$ below and check that it continues to tend to $1$.
For the sake of generality, we drop the $i,j$ indices, and consider a cooling probability restricted by a $k$-independent factor $A$.

With this adjustment, we may recast Eq.\ref{eq:cooling_k_c} when $A<<1$ as
\begin{equation}
1-p_c^{k_c}<\prod_{k,\frac{|E-\epsilon_k|}{\gamma_k}<1}\left[\left(1-\frac{A\pi^2}{4}\right)+\frac{A\pi^4}{48}\frac{(E-\epsilon_k)^2}{\gamma_k^2}\right].
\end{equation}
Then, taking the log and converting again to an integral, we obtain
\begin{equation}
\log(1-p_c^{k_c}) <\frac{1}{\alpha}\int_{\epsilon^-}^{\epsilon^+}\log\left[B+A'\frac{(E-\epsilon)^2}{\gamma(E)^2}\right]\frac{d\epsilon}{\gamma(\epsilon)},
\end{equation}
where $A'=\frac{A\pi^4}{48}\sim 2A$, and $B=1-\frac{A\pi^2}{4}<1$.
Next, setting $x=\frac{E-\epsilon}{\gamma(\epsilon)}$, and using the fact that for the LogSweep protocol $\gamma(\epsilon)\sim \epsilon$, we find
\begin{equation}
\log(1-p_c^{k_c}) < \frac{2}{\alpha}\int_{-1}^{+1}\log\left(B+A'x^2\right) dx.
\end{equation}
This may be evaluated by integrating by parts, giving
\begin{align}
\log(1-p_c^{k_c}) &< -\frac{2}{\alpha}\int_{-1}^{+1}\frac{x^2}{BA'^{-1} + x^2}\\
&=\frac{-4}{\alpha}\left[1-BA'^{-1}\tan^{-1}\left(A'B^{-1}\right)\right]\\
&\sim -\frac{4}{3\alpha}A'^2B^{-2}+O(A^{4}).
\end{align}
Using the optimal scaling $\alpha(K)=4\log^{-1}(K)$ we identified in Appendix~\ref{app:asymptotics}, this adjusts our bound in the cooling rate to
\begin{equation}
	p_c^{k_c}\gtrsim 1-K^{-\frac{1}{3}A'^2B^{-2}},
\end{equation}
which continues to tend to $1$ as $K\rightarrow\infty$, albiet at a rate reduced proportional to $A$.

This result requires some consideration in a large system --- if our coupling $\Gamma$ from a state $|E_i\rangle$ is spread over transitions to $J$ states $|E_j\rangle$, we have $A_{i,j}\sim J^{-1}$, and the probability of any transition being cooled can be found to be
\begin{equation}
\prod_{j}(1-p_{c,j}^{k_c})\sim e^{-\sum_{j=1}^J\frac{1}{3\alpha}A_j^2(1-A_j^2)}\sim e^{-\frac{1}{3\alpha J}}.
\end{equation}
This implies that we require $\alpha\sim J^{-1}$ in order to maintain a constant cooling rate, which in turn may require adjustments to the optimal scaling identified in Appendix~\ref{app:asymptotics}.
As such adjustments are highly system-dependent, we do not investigate them further here.

\section{Effect of banding on QDC protocols}

\begin{figure}[h] 
	\includegraphics[width=1\columnwidth]{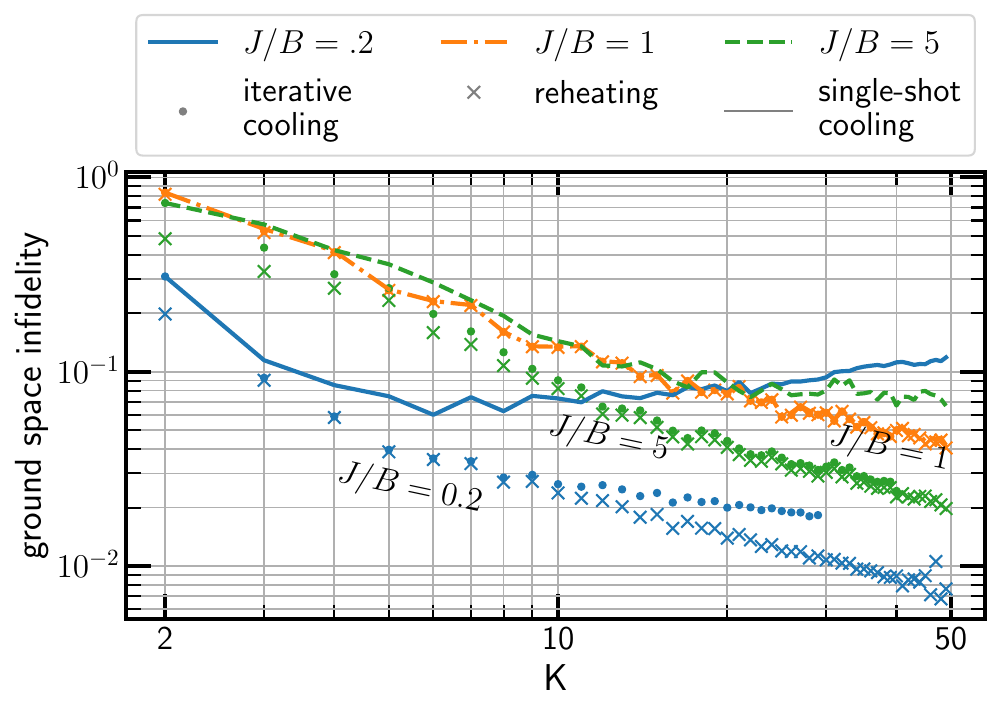}
	\caption{
		Difference between cooling by applying the a single LogSweep protocol with gradation number $K$ (round markers), and iterating LogSweep for all $g_l=2,...,K$ (solid lines).
		The iterative and reheating data are the same as in Fig.~\ref{fig:K_infidelity}, the same context and simulation techniques apply.
	} 
	\label{fig:K_infidelity-2}
\end{figure}

In this appendix we demonstrate the effect of banding on single sweeps of the LogSweep protocol.
In Fig.~\ref{fig:K_infidelity-2}, we plot the infidelity of a single shot of the LogSweep protocol with gradation number $K$ acting on the maximally-mixed state, as a function of $K$ (triangular markers). 
We see that in the critical case, the system continues to tend to the ground state polynomially in $K$.
However, for the TFIM chain in the weakly- and strongly-coupled phases, we find that the protocol fails to converge as a function of $K$, due to the banding issue described above.
This lack of convergence is rectified in the series marked by dots (same data as in Fig.~\ref{fig:K_infidelity}) by repeating the LogSweep protocol as a function of $K$.
We note that the failure in the strong-coupling case is not of the same degree as in the weak-coupling case, which we ascribe to the fact that the banding is not as strongly pronounced in Fig.~\ref{fig:banding}, and so the result has not yet presented itself.

\section{Numerical methods} \label{app:methods}

In this appendix we report the methods used to simulate QDC protocols on many-qubit systems.
The python code is packaged and available on github, at \url{https://github.com/aQaLeiden/QuantumDigitalCooling} \cite{QDCrepo}, and makes cirq \cite{cirq} to build and simulate the required quantum circuits.
All the data reported in this paper, and more simulations results that are left out in the interest of space and clarity, can be found in the same repository.

The simulations for the BangBang protocol energy validation Fig.~\ref{fig:energyguess}, as well as the study of LogSweep performance with increasing $K$ (Figures \ref{fig:K_infidelity} and \ref{fig:K_infidelity-2}) were performed using cirq density matrix simulator.
This stores the system's state in a density matrix, to which are applied sparse unitaries (representing the circuit's unitary gates) and the eventual quantum channel representing the reset gate.
For BangBang, as prescribed by the protocol, the unitary circuit applied before each fridge reset gate is defined by second-order Trotter expansion of the coupled system-fridge Hamiltonian with a single step (Trotter number $M=1$). 
This corresponds to Eq.~\ref{eq:trotter} where also $e^{-i H_\tS}$ is substituted by its $M=1$ second-order Trotter expansion.

To push to a larger number of qubits the results on scaling of both protocols, data for in Figures \ref{fig:bangbang} and \ref{fig:LSscaling} were generated with cirq state vector simulator.
The non-unitary reset required by QDC protocols and the initially-mixed state used to benchmark cooling cannot be represented deterministically in state vector simulations.
These are instead implemented by random sampling.
Each sample is constructed by choosing at random an initial computational basis state (these are enough to sample the maximally mixed state, because of the density matrix equivalence class).
For each non-unitary reset gate, the outcome is sampled probabilistically.
This process is repeated for 100 samples for each data point. 
The mean value of the quantity of interest is plotted, together with an interval representing the standard deviation of the mean.

In both the density matrix and the wavefunction simuations performed with cirq, the numerical error causes the final state to be often non normalized.
In the worst cases, the deviations from unit L2 norm (for wavefunctions) and trace norm (for density matrices) are up to few parts per thousand and percent respectively.
This is attributed to the large number of short-time Trotter steps required to produce LogSweep data in Fig.~\ref{fig:K_infidelity}, which translate to sparse unitaries with small entries and a building up of numerical error.
As all operations are linear, the first-order error can be dealt with by forcing normalization on the final state.
This technique was used for results reported in figures \ref{fig:K_infidelity}, \ref{fig:LSscaling} and \ref{fig:K_infidelity-2}

The numerical error of Trotterized sparse-unitary simulations still is too large to show the final state occupations in Fig.~\ref{fig:banding}, which range over more than 18 orders of magnitude.
For this reason, these simulations were performed by constructing the continuous evolution operator $e^{-i (H_\tS + H_\tF + H_\tC) t}$ for each unitary evolution step.
These results were validated by comparing with the Trotterized approach the results for large occupations and the energy expectation values (which are less sensitive to the numerical error).

\end{document}